\documentclass[aps,nofootinbib]{revtex4}
\usepackage{hetstyle}
\usepackage{graphicx}
\usepackage{psfrag}

\newcommand{\ra}{\rightarrow}

\newcommand{\sixj}[6]{\left\{ \begin{matrix} #1&#2&#3\\ #4&#5&#6
\end{matrix} \right\} }
\newcommand{\threej}[6]{\left( \begin{matrix} #1&#2&#3\\ #4&#5&#6
\end{matrix} \right) }
\newcommand{\cg}[3]{( #1\,;\,#2 \,|\, #3 )}
\newcommand{\weck}[3]{( #1\,||\,#2 \,||\, #3 )}

\newcommand{\bma}{\begin{matrix}}
\newcommand{\ema}{\end{matrix}}

\newcommand{\ad}[2]{a_{#1#2}^{\dagger}}
\newcommand{\ot}{\otimes}
\renewcommand{\a}[2]{a_{#1#2}}
\renewcommand{\l}{\left}
\renewcommand{\r}{\right}

\begin{document}

\title{Spin-orbit interaction in quantum dots in the presence of exchange correlations}
\author{Hakan E. T\"ureci and Y. Alhassid}
\affiliation{Center for Theoretical Physics, Sloane Physics Laboratory, Yale University, New Haven, CT 06520}
\setlength{\unitlength}{1mm}

\begin{abstract}
We discuss the problem of spin-orbit interaction in a 2D chaotic or diffusive quantum dot in the presence of exchange correlations. Spin-orbit scattering breaks spin rotation invariance, and in the crossover regime between different symmetries of the spin-orbit coupling, the problem has no closed solution. A conventional choice of a many-particle basis in a numerical diagonalization is the set of Slater determinants built from the single-particle eigenstates of the one-body Hamiltonian (including the spin-orbit terms). We develop a different approach based on the use of a good-spin many-particle basis that is composed of the eigenstates of the universal Hamiltonian in the absence of spin-orbit scattering. We introduce a complete labelling of this good-spin basis and use angular momentum algebra to calculate in closed form the matrix elements of the spin-orbit interaction in this basis. Spin properties, such as the ground-state spin distribution and the spin excitation function, are easily calculated in this basis.
\end{abstract}


\maketitle

\section{Introduction}
  
The statistical properties of the single-particle levels and wave functions in a quantum dot whose single-electron dynamics is chaotic can be described by random matrix theory (RMT)~\cite{alhassid00,guhr98}. More precisely, the RMT description holds for $\sim g_T$ single-particle levels around the Fermi energy, when $g_T \sim \sqrt{n}$ is the Thouless conductance of a dot with $n$ electrons.  In a dot that is strongly coupled to external leads (i.e., an open dot), the RMT description of non-interacting electrons can be used to derive the mesoscopic fluctuations of the conductance.   

However, if the dot is weakly coupled to the leads (i.e., an almost-isolated dot), the number of electrons in the dot is quantized and it is necessary to take into account  electron-electron interactions. The simplest model in that case is the constant interaction (CI) model, in which a classical charging energy $e^2n^2/2C$ ($C$ is the capacitance of the dot) is added to the single-particle Hamiltonian. This charging energy term is responsible for the observed Coulomb blockade peaks in the conductance as a gate voltage on the dot is varied. In a chaotic dot, the heights of these conductance peaks exhibit mesoscopic fluctuations. The conductance peak height distributions in the absence and presence of a  time-reversal symmetry-breaking orbital magnetic field were derived using RMT \cite{jalabert92} and were found to be in overall agreement with the experiments \cite{folk96,chang96}. However, experimental studies of the peak spacing statistics \cite{sivan96,simmel97,patel98a,luscher01} and the temperature dependence of the peak height statistics \cite{patel98b} suggested that it is necessary to include interaction effects beyond the CI model. 

A systematic approach to correlations in chaotic dots is based on a screened Coulomb interaction. The randomness of the single-particle wave functions induces randomness in the interaction matrix elements when the latter are calculated in the single-particle eigenstates~\cite{alhassid05}. It is possible to separate these matrix elements into an average part and a fluctuating part. The fluctuating part is of the order $1/g_T$~\cite{blanter97} and can be ignored in the limit of large $g_T$. The average part survives in this limit and leads to the so-called universal Hamiltonian~\cite{kurland00,aleiner02}. While there exists a strong-coupling phase whose fluctuation properties differ from those of the universal Hamiltonian~\cite{murthy03,murthy04}, in this work we assume the weak-coupling phase in which the universal Hamiltonian description is valid. The interaction part of the universal Hamiltonian contains, in addition to the charging energy term (which characterizes the CI model), an exchange interaction proportional to $\hat{\bm{S}}^2$, where $\hat{\bm{S}}$ is the total spin of the dot, and, in the absence of a time-reversal symmetry breaking orbital magnetic field, a Cooper channel term. The latter is repulsive in quantum dots and can be ignored. 

In the absence of spin-orbit interaction, the random matrix ensembles describing the single-electron Hamiltonian are the Gaussian orthogonal ensemble (GOE) and the Gaussian unitary ensemble (GUE) in the absence or presence of a time-reversal symmetry breaking orbital magnetic field, respectively. The exchange interaction modifies the conductance~\cite{master,exchange} and affects significantly the statistical fluctuations of both the conductance peak spacings and peak heights. Using a realistic value of the exchange interaction constant ($J_s=0.3 \Delta$), good agreement was found~\cite{exchange} between theory and the experiments of Ref.~\cite{patel98a} for the temperature dependence of the peak spacing fluctuations, explaining previous discrepancies with the predictions of the CI plus RMT model.  Discrepancies between the CI plus RMT model and the experiment of Ref.~\cite{patel98b} for the peak height statistics at low and intermediate temperatures ($k T \leq 0.6 \Delta$) are also explained by exchange correlations~\cite{exchange,usaj03}.

The presence of spin-orbit coupling in a 2D quantum dot was shown to lead to new RMT symmetry limits~\cite{aleiner01}. In particular, there are two spin-orbit terms that modify the vector potential of the orbital field: a term ${\bm a_\perp}$ that is proportional to the spin component $s_z$ (perpendicular to the plane of the dot), and a spin-flip term ${\bm a_\parallel}$. In the absence of a field parallel to the plane of the dot, these two terms lead to two new symmetry classes, both for conserved and broken time-reversal symmetry (i.e., in the absence and presence of an orbital magnetic field). 
The predictions of the modified single-electron theory were confirmed in experimental studies~\cite{zumbuhl02} of large open dots, in which electron-electron interactions can be ignored (spin-orbit effects are enhanced in large dots). 

The universal Hamiltonian of an almost-isolated dot and in particular the exchange interaction term are modified in these new symmetry limits of the spin-orbit interaction~\cite{so-universal}. The mesoscopic fluctuations of the conductance peak heights and peak spacings in these limits were studied and found to exhibit an interesting interplay between the spin-orbit and exchange interactions. 

However, in the crossover regimes between different symmetries, the Hamiltonian of the dot is no longer universal. Such crossovers describe many-body problems that cannot be solved in closed form. For a numerical diagonalization of the Hamiltonian, it is necessary to calculate its matrix elements in a many-particle basis. A conventional choice of a many-particle basis is obtained by finding the single-particle eigenstates of the one-body part of the Hamiltonian and constructing a complete set of Slater determinants that correspond to the occupied single-particle levels. The one-body Hamiltonian is diagonal in this basis, and the main part of the calculation of the Hamiltonian matrix elements is the computation of the two-body interaction matrix elements in this basis. Such a basis was used in Ref.~\cite{gorokhov03}, which studied the combined effects of spin-orbit and exchange interactions on the $g$-factor statistics in metallic nanoparticles~\cite{vondelft01}. In such 3D nanoparticles there are no intermediate symmetries and the only relevant crossover (in the absence of an orbital magnetic field) is between the GOE and the Gaussian symplectic ensemble (GSE). A basis of Slater determinants was also used in studying the ground-state magnetization properties of a disordered quantum dot in the presence of exchange correlations and fluctuations of the off-diagonal interaction matrix elements~\cite{jacquod01}.

Here we consider another choice of the many-particle basis. We use a good total spin basis composed of the eigenstates of the universal Hamiltonian in the absence of spin-orbit interaction. There are several advantages in using such a basis. First, the calculation of the matrix elements of $H$ requires the evaluation of the matrix elements of the spin-orbit terms, which are one-body operators, in contrast to the conventional basis in which it is necessary to calculate the matrix elements of the interaction, a two-body operator. Second, the spin-orbit terms can be expressed as tensor operators under spin rotations, and we can take advantage of angular momentum algebra in the calculation of their matrix elements. In particular, matrix elements of both the ${\bf a}_\perp$ and ${\bf a_\parallel}$ spin-orbit terms can be calculated from the same reduced matrix elements of certain tensor operators.  Third, the eigenstates of the universal Hamiltonian can be organized in ascending energy and constitute a suitable basis for truncated calculations. In particular, an attractive exchange interaction brings down in energy high-spin states that in absence of exchange lie higher in energy.  Finally, the good-spin basis is very convenient for calculating spin properties of the dot such as the ground-state spin distribution and the spin excitation function.

The current work introduces the good-spin basis formalism. This approach is useful in various problems involving exchange correlations. For example, it would be interesting to study signatures of quantum critical fluctuations in the spin excitation function that are predicted in Ref.~\cite{murthy04a} for spin-orbit crossovers in the presence of correlations. We emphasize, however, that our formalism is quite general and is not limited to spin-orbit scattering. It can, for example, be applied to a quantum dot in which the electrons interact ferromagnetically between themselves and antiferromagnetically with a spin-$1/2$ Kondo impurity~\cite{murthy05}. In this problem the local spin of the dot plays the role of the one-body operator (instead of the spin-orbit terms).

\section{Hamiltonian}

In the absence of spin-orbit interaction, the Hamiltonian of the quantum dot in the limit of large Thouless conductance $g_T$ is given by the so-called universal Hamiltonian 
\be\label{universal-H}
\hat{H}= \sum_{\mu \sigma}\epsilon_\mu a^\dagger_{\mu \sigma} a_{\mu \sigma}+ \frac{e^2}{2C} \hat{n}^2 - J_s \hat{\bm{S}}^2 \;.
\ee
Here $a^\dagger_{\mu \sigma}$ is the creation operator of an electron in an orbital level $\mu$ and spin $\sigma/2$ ($\sigma=\pm 1$), and $\epsilon_\mu$ are spin-degenerate single-particle levels. The last term on the r.h.s. of (\ref{universal-H}) is an exchange interaction with $\hat{\bm{S}} = \frac{1}{2} \sum_{\mu,\sigma,\sigma'} a^\dagger_{\mu \sigma}\bm{\sigma}_{\sigma \sigma'} a_{\mu \sigma'}$ being the total spin operator of the quantum dot ($\bm\sigma$ are Pauli matrices) and $J_s$ is the exchange interaction constant. 
The single-particle levels $\epsilon_\mu$ are spin-degenerate and correspond to the eigenvalues of a single-particle Hamiltonian $\hat h = (-i \hbar{\bf \grad} - e {\bf A})^2/2m + V({\bf r})$, where ${\bf A} = B_3 ({\vec e}_3 \times {\bf r})/2c$ is a vector potential describing an orbital magnetic field $B_3$ perpendicular to the plane of the dot and $V({\bf r})$ is a confining potential. In a chaotic dot, these single-particle levels  follow RMT statistics, i.e., GOE (GUE) in the absence (presence) of an orbital magnetic field.

In the presence of spin-orbit interaction, the single-particle Hamiltonian is modified. Performing a gauge transformation to a locally-rotated spin frame~\cite{aleiner01}, the single-particle Hamiltonian is found to have the form $\hat h = (-i \hbar{\bf \grad} - e {\bf A} -{\bf a}_\perp - {\bf a}_\parallel)^2/2m + V({\bf r})$, where
\be\label{so-a}
{\bf a}_\perp = {{\bf e}_3\times {\bf r} \over 4 \lambda_1 \lambda_2} \hat \sigma_3 \;;\;\;\;
{\bf a}_\parallel = {{\bf e}_3\times {\bf r} \over 6 \lambda_1 \lambda_2} \left( {x_1 \over \lambda_1} \hat \sigma_1 + {x_2 \over \lambda_2} \hat \sigma_2 \right) 
\ee
are two effective vector potentials.
The ${\bf a}_\perp$ term describes an effective spin-orbit magnetic field that has opposite directions for spin up and spin down electrons, while ${\bf a}_\parallel$ describes a spin-flip term. $\lambda_1,\lambda_2$ are the spin-orbit scattering lengths for electrons moving along the principal crystallographic directions ${\bf e}_1, {\bf e}_2$ in the plane of the dot.

The many-particle Hamiltonian (\ref{universal-H}) is modified to include the spin-orbit Hamiltonian $\hat{H}_{\rm so}$
\be\label{so-exchange}
\hat{H}= \sum_{\mu \sigma}\epsilon_\mu a^\dagger_{\mu \sigma} a_{\mu \sigma} + \frac{e^2}{2C} \hat{n}^2 - J_s \hat{\bm{S}}^2 + \hat{H}_{\rm so}\;.
\ee
In a chaotic dot, the spin-orbit Hamiltonian $\hat{H}_{\rm so}$ can be represented in terms of random matrices of the appropriate symmetry. We have
\be\label{spin-orbit}
H_{\rm so} = i \alpha _\perp \sum_{\mu \nu \sigma} \sigma a^\dagger_{\mu \sigma} \Gamma^\perp_{\mu \nu} a_{\nu \sigma} + \left( i \alpha_\parallel \sum_{\mu \nu} a^\dagger_{\mu +} \Gamma^\parallel_{\mu \nu} a_{\nu -} +h.c.\right) \;,
\ee
where $\Gamma^\parallel = \Gamma_1 - i \Gamma_2$ and $\Gamma^\perp$, $\Gamma_1$, $\Gamma_2$  are real antisymmetric random matrices, and $\alpha_\perp,\alpha_\parallel$ are coupling parameters associated with the ${\bf a}_\perp$ and ${\bf a}_\parallel$ spin-orbit terms, respectively. 

Here we work in the basis of single-particle eigenstates $\mu$ in the absence of spin-orbit interaction.  These are the eigenstates of a real symmetric matrix (GOE) or a complex hermitian matrix (GUE) in the absence or presence of an orbital magnetic field, respectively.  If the variance of an off-diagonal element is chosen as $a^2/2$, then the mean single-particle level spacing in the middle of the spectrum is $\Delta= \pi a/\sqrt{2N}$ ($N$ is the dimension of the random matrix). The random Gaussian matrices $\Gamma^\perp$, $\Gamma_1$ and $\Gamma_2$ are also chosen to have variance $a^2/2$ for their off-diagonal elements. 

The strength of the spin-orbit coupling is characterized by two dimensionless crossover parameters
\be 
x_\perp = \pi {\alpha_\perp {\rm rms}(\Gamma^\perp_{\mu\nu}) \over \Delta} = \alpha_\perp \sqrt{N} \;,
\ee
and
\be 
x_{\parallel} = \pi {\alpha_\parallel {\rm rms}(\Gamma^\parallel_{\mu\nu}) \over \Delta} = \alpha_\parallel \sqrt{2 N} \;.
\ee

In the actual physical application, these crossover parameters are determined by the two spin-orbit energy scales $\epsilon_\perp$ and $\epsilon_\parallel$  
\be
x_\perp^2 = {\epsilon_\perp \over \Delta} = \kappa g_T \left({{\cal A} \over \lambda_1 \lambda_2} \right)^2 \;;
\ee
\be
x_\parallel^2 = {\epsilon_\parallel \over \Delta} = \kappa' \left[ \left({L_1\over \lambda_1}\right)^2 + \left({L_2\over \lambda_2}\right)^2 \right] x_\perp^2  \;.
\ee
Here $L_{1,2}$ are the corresponding linear lengths of the dot, ${\cal A}\sim L_1 L_2$ is the dot's area, and $\kappa$, $\kappa'$ are constants of order unity.

\section{Good spin many-particle basis}

A numerical diagonalization of a many-particle Hamiltonian of the type (\ref{so-exchange}) requires a choice of a many-particle basis. In the conventional approach, the many-particle basis is chosen to be the set of Slater determinants of the single-particle eigenstates of the full one-body Hamiltonian $\sum_{\mu \sigma}\epsilon_\mu a^\dagger_{\mu \sigma} a_{\mu \sigma}+\hat H_{\rm so}$. However, here  
we use a good-spin basis determined by the eigenstates of the universal Hamiltonian (\ref{universal-H}) (in the absence of spin-orbit interaction). To find such a basis, we note that the occupation number operator $\hat{n}_\lambda = \hat{n}_{\lambda+} + \hat{n}_{\lambda-}$ of any single particle  orbital $\lambda$ is a scalar under rotations in spin space and therefore commutes with the total spin operator, $[\hat{n}_\lambda,\hat{\bm{S}}]=0$. We can then characterize the eigenstates of the universal Hamiltonian by the orbital occupations $\bm{n}=\{ n_{\lambda} \}$ ($n_\lambda=0,1,2$), total spin quantum number $S$ and the spin projection $S_z=M$. Additional quantum numbers $\gamma$ are needed to distinguish between states with the same orbital occupations $\bm{n}$ and total spin $S$. 

The sequence of occupations  $\bm{n}=\{ n_{\lambda} \}$ can be alternatively specified by $\{\lambda_i\} \{\tilde{\lambda}_i\}$, where $\lambda_i$ ($\tilde{\lambda}_i$) denote the corresponding numbers of singly-occupied (doubly-occupied) levels.  The many-particle basis is then denoted by
 \be
 \bra{ \bm{n} \gamma S M} = 
\bra{ \{\lambda_i\} \{\tilde{\lambda}_i\} \gamma S M} \;.
\ee
 
\subsection{Labelling of the many-body states}

A complete labelling of the good-spin states requires the specification of the quantum numbers $\gamma$. These quantum numbers can be described by a tree graph, as we explain in the following. 

Consider a set of orbital occupations $\bm{n}$ with $q$ singly occupied levels $\lambda_1,\lambda_2,\ldots,\lambda_q$ where $\lambda_q > \cdots \lambda_2 > \lambda_1$. They describe a manifold of $2^q$ Slater determinants 
\be
\ad{\lambda_1}{\sigma_1}\ad{\lambda_2}{\sigma_2}\ldots \ad{\lambda_q}{\sigma_q} \bra{\{\tilde{\lambda}_i\}, S=0}
\label{eqslater1}
\ee
with $\sigma_i=\pm 1$ ($i=1,\ldots,q$). Since the doubly-occupied levels $\tilde\lambda_i$ have spin zero, the construction of good-spin basis within this manifold is equivalent to the problem of coupling $q$ spin-$1/2$ particles to total spin $S$. The total spin $S$ can then take the values $q/2,q/2-1,...,q/2-[q/2]$. A unique labelling of the states requires us to choose a convention for the coupling sequence of the spin-$1/2$ particles. We assume that the spins are coupled in the ascending order of their orbital label
\be
\lambda_1 \otimes \lambda_2 \otimes \cdots \otimes\lambda_q \equiv \l( \left( \cdots \left(  \left( \lambda_1 \otimes \lambda_2 \right) \otimes \lambda_3 \right) \otimes  \cdots \r) \otimes \lambda_q \right) \;.
\ee

\begin{figure}[!hbt]
\centering
\psfrag{l1}{\Large $\lambda_1$}
\psfrag{l2}{\Large $\lambda_2$}
\psfrag{l3}{\Large $\lambda_3$}
\psfrag{l4}{\Large $\lambda_4$}
\psfrag{l5}{\Large $\lambda_5$}
\includegraphics[width=0.6\linewidth]{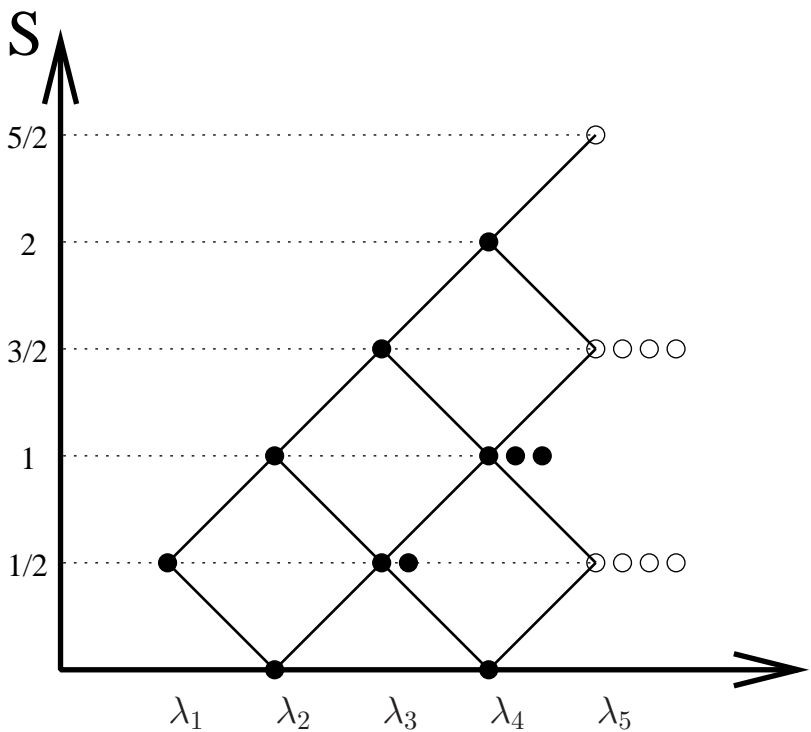}
\caption{An example of a tree graph representing the normal-coupling scheme 
$\lambda_1 \otimes \lambda_2 \otimes \lambda_3 \otimes \lambda_4 \otimes\lambda_5$ for $\lambda_5 > \cdots \lambda_2 > \lambda_1$ singly-occupied levels. The total spin $S$ of each state is indicated on the vertical axis. The $d_q(S)$ states with the same total spin $S$ are distinguished by the different paths $\gamma$ of intermediate spins. The internal points are shown by solid circles, whereas the terminal points are shown by open circles.}
\label{figStree1}
\end{figure}

Fig.~\ref{figStree1} illustrates this coupling scheme in a diagram showing the total spin $S$ versus the number of singly-occupied levels $q$. The diagram has a tree structure. We call the first single-particle orbital state $\lambda_1$ the {\em root} of the tree. The root bifurcates into two points with total spin $S=0$ and $S=1$. The point $S=1$ bifurcates into two more points with $S=1/2$ and $S=3/2$, while the point $S=0$ can only connect to an $S=1/2$ state. This process continues with each point bifurcating into two (except for the $S=0$ points) until we reach the states with $q$ singly-occupied levels. The points on the tree with $q$ singly-occupied levels are called {\em terminal points} (open circles) while all other points between the root and the terminal points are called {\em internal points} (solid circles). We observe that in general there are multiple states with the same spin $S$ (and orbital occupations $\bm{n}$). The multiplicity is given by the combinatorial number
\be
d_q(S)= \left( \bma q \\ S+\frac{q}{2} \ema \right) - \left( \bma q \\ S+1+\frac{q}{2} \ema \right)  \;.
\ee
These states are degenerate eigenstates of the universal Hamiltonian and we distinguish them by additional quantum numbers $\gamma$. A convenient choice is the path used to generate the state in the coupling sequence, i.e., the spin sequence of the internal points, 
\be
(\gamma, S) \equiv (S_1=1/2,S_2,S_3,\ldots,S_{q-1},S_q=S) \;.
\ee
This set of quantum numbers corresponds to a complete set of commuting observables $(\bm{S}^2_{12},\bm{S}^2_{123},\ldots,\bm{S}^2_{12\ldots(q-1)},\bm{S}^2)$, where  $\bm{S}_{12\ldots j}$ are intermediate spin operators.\footnote{ We use $S_j$ to denote the spin of the coupled spin angular momentum ${\bm S}_{12\ldots j}= {\bm S}_1 + {\bm S}_2 +\ldots + {\bm S}_j$.}

\section{The transformation from a normal tree to a canonical tree}
\label{sectchanofbas}

We call the tree introduced in the previous section a {\em normal} tree. The orbital levels  $\lambda_1, \lambda_2,\cdots,\lambda_q$ are ordered in ascending order and the normal-tree states are given by
\be
\bra{\{\lambda_i\} \gamma S M}_\lambda = \left| \left( \left(  \left( \lambda_1 \otimes \lambda_2 \right)_{S_2} \otimes \lambda_3 \right)_{S_3} \otimes \cdots \otimes \lambda_q \right)_S M \right\rangle \;,
\ee
where for simplicity we have omitted the labels $\tilde\lambda_i$ of the doubly-occupied levels.

However, calculations of the matrix elements of $a^\dagger_{\mu\sigma} a_{\nu\sigma'}$ (which appear in the spin-orbit interaction) are best carried out in the {\em canonical} basis, i.e., the $\mu$-tree (or $\nu$-tree). This is the coupling scheme where the last coupled spin is that of the orbital $\mu$ ($\nu$) corresponding to the orbital label of the creation (annihilation) operator. This last orbital $\mu$ can generally be out of order, as is the case for $\lambda_j=\mu$ and $j<q$. 

The normal- and canonical-tree bases are related by a unitary transformation, which, as we now show, can be expressed in terms of the Wigner 6j symbols.  The transformation between the normal and canonical bases can be constructed by successive applications of two basic operations: {\em commutation} and {\em association}.  The commutation of two spins $j_1,j_2$ coupled to spin $j_{12}$ is simply a phase
\be\label{commutation}
(j_1 \otimes j_2)_{j_{12}} \stackrel{(-1)^\phi}{\ra} (j_2 \otimes j_1)_{j_{12}} \;,
\ee
where $\phi=j_1+j_2-j_{12}$, as is easily verified by the corresponding symmetry of the Clebsch-Gordan coefficients. The operation $\mcal{R}$ of association among three spins $j_1$,$j_2$,$j_3$ coupled to total spin $J$ 
\be\label{association}
\left( \left( j_1 \otimes j_2 \right)_{j_{12}} \otimes j_3 \right)_J \stackrel{\mcal{R}}{\ra} \left( j_1 \otimes \left( j_2  \otimes j_3 \right)_{j_{23}} \right)_J    
\ee
is given in terms of the Wigner 6j symbol\footnote{Relation (\ref{association1}) essentially constitutes the definition of the 6j symbol.}
\be\label{association1}
{\cal R}_{j_{12},j_{23}}=(-1)^{j_1+j_2+j_3+J}\sqrt{(2j_{12}+1)(2j_{23}+1)} \sixj{j_1}{j_2}{j_{12}}{j_3}{J}{j_{23}} \;.
\ee
The matrix $\mcal{R}$, defined for given spins $j_1,j_2,j_3$ and $J$, and labelled by the intermediate spin values $j_{12}$ and $j_{23}$,  is a real orthogonal transformation.{\footnote{The inverse transformation   $j_1(j_2j_3) \ra (j_1j_2)j_3$ is simply obtained by inverting the sign of the phase in (\ref{association1}).}

We consider the unitary transformation which takes us from the normal basis to a canonical basis where the last coupled spin is $\mu=\lambda_j$ ($j < q$), the rest of the spins being coupled in their normal order. Such a transformation can be obtained by successive applications of the {\em exchange} operation $R$ which exchanges the coupling order of two neighboring spins 
\be
\l( \l( \Lambda_{S_{i-1}} \ot \lambda_i \r)_{S_i} \ot  \lambda_{i+1} \r)_{S_{i+1}} \stackrel{R}{\ra}  \l( \l( \Lambda_{S_{i-1}} \ot \lambda_{i+1} \r)_{S_i'} \ot  \lambda_{i} \r)_{S_{i+1}} \;,
\ee
where $\Lambda_{S_{i-1}}$ denotes a sequence of orbital labels coupled to spin $S_{i-1}$.  The exchange $R$ can be obtained by the following successive application of commutation, association and commutation\footnote{There are other sequences of operations that lead to the same exchange, but they can be shown to lead to the same matrix through the use of Wigner 6j identities. The sequence given in Eq.~(\ref{exchange}) is the ``shortest path."}
\begin{eqnarray}\label{exchange}
\l( \l( \Lambda_{S_{i-1}} \lambda_i \r)_{S_i}\lambda_{i+1} \r)_{S_{i+1}} \stackrel{(-1)^\phi}{\ra} \l( \lambda_{i+1}\l( \Lambda_{S_{i-1}} \lambda_i \r)_{S_i} \r)_{S_{i+1}}  \stackrel{\mcal{R}^{-1}}{\ra}\l( \l( \lambda_{i+1}\Lambda_{S_{i-1}} \r)_{S_i'}\lambda_i \r)_{S_{i+1}} \\ \nonumber \stackrel{(-1)^\phi}{\ra} \l( \l( \Lambda_{S_{i-1}}\lambda_{i+1} \r)_{S_i'}\lambda_i \r)_{S_{i+1}} \;.
\end{eqnarray}
Using (\ref{commutation}) and (\ref{association1}), we find
\be
R_{S_i,S_i'} \equiv R_{S_i,S_i'}(S_{i-1},S_{i+1}) = (-1)^{S_i-2S_{i+1}-S_i'} \sqrt{(2S_i'+1)(2S_i+1)}\sixj{\frac{1}{2}}{S_{i-1}}{S_i'}{\frac{1}{2}}{S_{i+1}}{S_i} \;,
\label{eqpushout1}
\ee
where $R(S_{i-1},S_{i+1})$ is a $2\times 2$ matrix which depends parametrically on two spins $S_{i-1},S_{i+1}$.  As for any exchange operator, $R^2=1$.

The transformation from the normal basis to a given canonical basis can be achieved by a succession of exchange operators. It is important however to keep in mind that the many-particle states are defined in terms of the product of fermionic creation operators and not simply spins. The series of exchanges only reorders the labels of the operators, but we also want to move the corresponding creation operator itself to the top of the string of operators. This results in an overall sign which is determined by the parity number (even or odd) of exchanges. To take into account this fermionic sign, we simply redefine $R$ in (\ref{eqpushout1}) 
\begin{equation}
R_{S_i,S_i'} \equiv R_{S_i,S_i'}(S_{i-1},S_{i+1}) = (-1)^{S_i-2S_{i+1}-S_i'+1} \sqrt{(2S_i'+1)(2S_i+1)}\sixj{\frac{1}{2}}{S_{i-1}}{S_i'}{\frac{1}{2}}{S_{i+1}}{S_i} \;.
\label{eqpushout2}
\end{equation}

\section{Calculation of Matrix elements}

\subsection{The spin-orbit Hamiltonian in terms of irreducible tensor operators}

The use of irreducible tensor operators simplifies the calculation of the matrix elements through the application of the Wigner-Eckart theorem. Since both $\ad{\lambda}{\sigma}$ and $\tilde{a}_{\lambda\sigma}\equiv (-1)^{(1+\sigma)/2}a_{\lambda-\sigma}$ transform as irreducible tensor operators of rank $1/2$ under spin rotations, we can form the following irreducible tensor-operators from their bilinear products
\be
(A_{\mu\nu})^{\Sigma}_{m} \equiv \left( \ad{\mu}{} \otimes \tilde{a}_{\nu}{} \right)^{\Sigma}_{m} = \sum_{\sigma\sigma'} \cg{1/2,\sigma/2}{1/2,\sigma'/2}{\Sigma,m} \ad{\mu}{\sigma}\tilde{a}_{\nu\sigma'} \;,
\label{eqirrTOgen}
\ee
where $\Sigma=0$, $m=0$ or $\Sigma=1$, $m=0,\pm 1$. 
Eq.~(\ref{eqirrTOgen}) reads
\be\label{tensors}
(A_{\mu\nu})^{0}_{0} = \frac{1}{\sqrt{2}} \l( \ad{\mu}{+}\a{\nu}{+} + \ad{\mu}{-}\a{\nu}{-} \r)
\ee
and
\be\label{tensors1}
(A_{\mu\nu})^{1}_{0} = \frac{1}{\sqrt{2}} \l( \ad{\mu}{+}\a{\nu}{+} - \ad{\mu}{-}\a{\nu}{-} \r)\;;\;\;\; (A_{\mu\nu})^{1}_{1} = - \ad{\mu}{+}\a{\nu}{-}\;;\;\;\;
(A_{\mu\nu})^{1}_{-1} =  \ad{\mu}{-}\a{\nu}{+}\;.
\ee
Using relations (\ref{tensors}) and (\ref{tensors1}), we can rewrite the spin-orbit interaction terms (\ref{spin-orbit}) as
\be
H_{\rm so} = i \sqrt{2}\alpha_\perp \sum_{\mu \nu} \Gamma^\perp_{\mu\nu} \l( A_{\mu\nu}\r)^1_0 - \left( i \alpha_\parallel \sum_{\mu \nu} \Gamma^\parallel_{\mu \nu} \l( A_{\mu\nu}\r)^1_1  +h.c.\right) \;.
\ee
Both the $\bm{a}_\perp$ and $\bm{a}_\parallel$ spin-orbit terms are components of a $\Sigma=1$ tensor and thus can change the total spin of the dot by $\Delta S =0, \pm 1$. However, the $\bm{a}_\perp$ spin-orbit term is the $m=0$ component of this tensor, thus conserving $S_z$, while the spin-flip $\bm{a}_\parallel$ term is composed of the $m=\pm 1$ components, thus changing $M$ by $\pm 1$.  

\subsection{Calculation of the reduced matrix elements}

According to the Wigner-Eckart theorem, the matrix elements of a tensor operator in a good angular momentum basis can be factorized into a reduced matrix element and a Wigner 3j symbol
\be\label{eqweckthe1}
\ket{\gamma' S'M'}(A_{\mu\nu})^\Sigma_m \bra{\gamma SM}
=(-1)^{S'-M'}\threej{S'}{\Sigma}{S}{-M'}{m}{M} \weck{\gamma' S'}{A_{\mu\nu}^\Sigma}{\gamma S} \;.
\ee
 The reduced matrix element $\weck{\gamma' S'}{A_{\mu\nu}^\Sigma}{\gamma S}$ is independent of the magnetic quantum numbers $M,M'$ and $m$. Once this reduced matrix element of  $A_{\mu\nu}^\Sigma$ is calculated, we can easily obtained the matrix elements of both the $\bm{a}_\perp$ and $\bm{a}_\parallel$ spin-orbit terms, using (\ref{eqweckthe1}) with $m=0$ and $m=\pm 1$, respectively (for $\Sigma=1$). 

The reduced matrix elements of the spin-orbit operators $A_{\mu\nu}^\Sigma$ can be grouped into four different classes depending on the initial occupations of the orbitals $\mu$ ($0$ or $1$) and $\nu$ ($1$ or $2$):
\begin{eqnarray}
(i)\;\;\; n_\nu = 1 &\ra& n_\nu' = 0 \;,\;\;\; n_\mu = 0 \ra n_\mu' = 1 \;; \nonumber \\
(ii)\;\;\; n_\nu =1 &\ra& n_\nu' = 0 \;,\;\;\; n_\mu=1 \ra n_\mu' = 2 \;; \nonumber \\
(iii)\;\;\; n_\nu =2 &\ra& n_\nu' = 1 \;,\;\;\; n_\mu=1 \ra n_\mu' = 2 \;; \nonumber \\
(iv)\;\;\; n_\nu = 2 &\ra& n_\nu' = 1 \;,\;\;\; n_\mu=0 \ra n_\mu' = 1 \;.
\end{eqnarray}
Although the operators $A_{\mu\nu}$ conserve the total number of particles 
$n$, they do not necessarily conserve $q$ on the tree graph.  
While cases (i) and (iii) conserve $q$, case (ii) is characterized by a
$q\ra (q-2)$ transition and case (iv) by a $q\ra (q+2)$ transition.  In the following, we need to consider separately these four classes of transitions.

Below we outline the calculation of the reduced matrix element for case (i) in detail, while for the remaining three cases we only present the results. 

\subsubsection{Transition from a singly-occupied level to an empty level}\label{case1}

The first case corresponds to the situation in which a particle is moved from a singly-occupied level $\nu$ to an empty level $\mu$. The evaluation of the matrix elements proceeds in two steps. First we consider the evaluation of the reduced matrix element $\weck{\bar{\gamma}' S'} {A_{\mu\nu}^\Sigma} {\bar{\gamma} S}_c$ in the canonical basis. Then we transform to the normal tree by application of appropriate orthogonal transformations introduced in Section~\ref{sectchanofbas}. Note that these transformations conserve the total spin, i.e. $\bar{S}=S$ and $\bar{S}'=S'$. 
 
Consider the reduced matrix element $\weck{\bar{\gamma}' S'} {A_{\mu\nu}^\Sigma} {\bar{\gamma} S}_c$ where the initial and final states (as well as the intermediate states) are in the canonical basis.\footnote{all canonical matrix elements will be denoted by the subscript c or else by the orbital label of the last coupled spin.} The action of $A_{\mu\nu}^{\Sigma=1}$ in the canonical basis is illustrated in Fig.~\ref{figcanop1}. Depending on $\bar{\gamma}$, the spin $S''$ of the intermediate state (after a particle is removed from the level $\nu$) can be either $S''=S+1/2$ or $S''=S-1/2$. We note that $S''$ is uniquely determined given the path $\bar{\gamma}$. In Appendix A we show that the canonical reduced matrix element is given by
\be
\weck{\bar{\gamma}' S'}{A_{\mu\nu}^\Sigma}{\bar{\gamma} S}_c = (-1)^{S'+S''-\Sigma+1/2} \sqrt{(2\Sigma+1)(2S+1)(2S'+1)} \sixj{S}{S'}{\Sigma}{1/2}{1/2}{S''} \;.
\label{eqredmatel1}
\ee
For $\Sigma=1$, the corresponding 6j symbol leads to the expected selection rule $S'=S,S \pm 1$.

\begin{figure}[!hbt]
\centering
\psfrag{l1}{\Large $\lambda_1$}
\psfrag{l2}{\Large $\lambda_2$}
\psfrag{l3}{\Large $\lambda_3$}
\psfrag{l4}{\Large $\lambda_4$}
\psfrag{l5}{\Large $\lambda_5$}
\psfrag{l6}{\Large $\lambda_6$}
\psfrag{l7}{\Large $\lambda_7$}
\psfrag{s4}{\Large $S''$}
\includegraphics[width=0.45\linewidth]{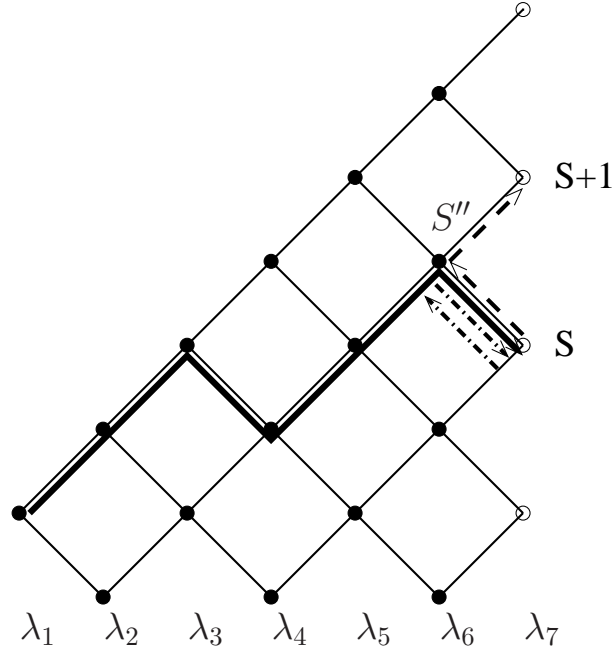}
\caption{An example of a $q=7$ tree, showing the action of $A_{\mu\nu}^{\Sigma=1}$ in the canonical basis. The thick solid line represents the initial state $\bar{\gamma}$ with $S=3/2$. Under the action of $A_{\mu\nu}^{\Sigma=1}$, this state is connected to states with $S'= S$ (dotted-dashed path) and $S'=S+1$ (dashed path). In the  case shown, the intermediate spin in the transition is $S''= S+ 1/2$ (determined by the initial path $\bar{\gamma}$).}   
\label{figcanop1}
\end{figure}
To evaluate the reduced matrix element in the normal basis
$\weck{\gamma' S'}{A_{\mu\nu}^\Sigma}{\gamma S}$,
we expand the initial and final normal states in terms of the respective canonical basis
\beq
\bra{\gamma SM} & = & \sum_{\bar{\gamma}} U^{\nu}_{\gamma\bar{\gamma}} \bra{\bar{\gamma} SM}_\nu \label{eqnormcan1} \\
\bra{\gamma' S'M'} & = & \sum_{\bar{\gamma}'} V^{\mu}_{\gamma'\bar{\gamma}'} \bra{\bar{\gamma}' S'M'}_\mu  \;,\label{eqnormcan2}
\eeq
where subscripts $\mu$ and $\nu$ refer to the orbital which is coupled last. Here the transformation matrices $U^\nu$ and $V^\mu$ are independent of the magnetic quantum numbers $M$ and $M'$. 
Then, using Eqs.~(\ref{eqredmatel1}), (\ref{eqnormcan1}) and (\ref{eqnormcan2}), we obtain
\beq\label{red1}
\weck{\gamma' S'}{A_{\mu\nu}^\Sigma}{\gamma S} =  (-1)^{S'+\Sigma +1/2} \sqrt{(2\Sigma+1)(2S+1)(2S'+1)} \nonumber\\
 \times  \sum_{\bar{\gamma},\bar{\gamma}'} (-1)^{S''}
U^{\nu}_{\gamma\bar{\gamma}} V^{\mu}_{\gamma'\bar{\gamma}'} \sixj{S}{S'}{\Sigma}{1/2}{1/2}{S''} \;.
\label{eqmatelred1}
\eeq
As is expected, no magnetic quantum numbers appear in (\ref{red1}).

To calculate the unitary transformation $U^\nu$ explicitly, assume the spin we move out is at position $i$ in the normal-ordered tree, i.e., $\lambda_i=\nu$. The transformation $U^\nu$ can then be viewed as a push-out operation 
\be
\bra{\lambda_1,\lambda_2,\ldots,\lambda_i=\nu,\ldots,\lambda_q; \gamma SM} \stackrel{U_{\gamma\bar{\gamma}}^\nu}{\ra} \bra{\lambda_1,\lambda_2,\ldots,\lambda_{i-1},\lambda_{i+1},\ldots,\lambda_q,\lambda_i=\nu; \bar{\gamma}SM}_{\nu} \;.
\ee
Defining $\gamma=(S_1=1/2,S_2,S_3,\ldots,S_q=S)$ and $\bar{\gamma}=(\bar{S}_1=1/2,\bar{S}_2,\bar{S}_3,\ldots,\bar{S}_q=S)$,
$U_{\gamma\bar{\gamma}}^\nu$ can be written in terms of the exchange operator defined in  Eq.~(\ref{eqpushout1})
\be
U^\nu_{\gamma\bar{\gamma}}=R_{S_i,\bar{S}_i}(S_{i-1},S_{i+1})\,R_{S_{i+1},
\bar{S}_{i+1}}(\bar{S}_{i},S_{i+2})\,R_{S_{i+2},\bar{S}_{i+2}}(\bar{S}_{i+1},S_{i+3})\,\cdots\, R_{S_{q-1},\bar{S}_{q-1}}(\bar{S}_{q-2},S_{q}=S) \;.
\label{eqpushoutU}
\ee
The matrices $R$ in (\ref{eqpushoutU}) are $2\times 2$  (because of the spin selection rules), and they depend parametrically on the neighboring spins. We should note that a special situation arises when the spin to be pushed out is at the root of the tree, i.e. $i=1$. In that case, a simple commutation is required to move the spin to the position $i=2$ and one can formally replace $R_{S_i,\bar{S}_i}(S_{i-1},S_{i+1})=(-1)^{S_2}$.

Similarly, $V^\mu$ can be viewed as a push-in operation
\be
\bra{\lambda_1,\lambda_2,\ldots,\lambda_q=\mu; \bar{\gamma}' S'M'}_\mu \stackrel{V_{\gamma'\bar{\gamma}'}^\mu}{\ra} \bra{\lambda_1,\lambda_2,\ldots,\lambda_{j}=\mu,\ldots,\lambda_q,; \gamma' S'M' } \;,
\ee
and is calculated to be
\begin{eqnarray}\label{eqpushinV}
V_{\gamma'\bar{\gamma}'}^\mu=R_{\bar{S}_{q-1}',S_{q-1}'}(\bar{S}_{q-2}',\bar{S}_q'=S')
\,R_{\bar{S}_{q-2}',S_{q-2}'}(\bar{S}_{q-3}',S_{q-1}')\,R_{\bar{S}_{q-3}',S_{q-3}'}
(\bar{S}_{q-4}',S_{q-2}') \nonumber \\ \cdots\,R_{\bar{S}_{j}',S_{j}'}(\bar{S}_{j-1}',S_{j+1}') \;.
\end{eqnarray}
Eqs.~(\ref{eqpushoutU}) and (\ref{eqpushinV}) allow us to represent the unitary push-in and push-out operations (which are of dimension $d_q(S)\times d_q(S)$) in terms of a string of scalar operations. 
\begin{figure}[!hbt]
\centering
\includegraphics[width=\linewidth]{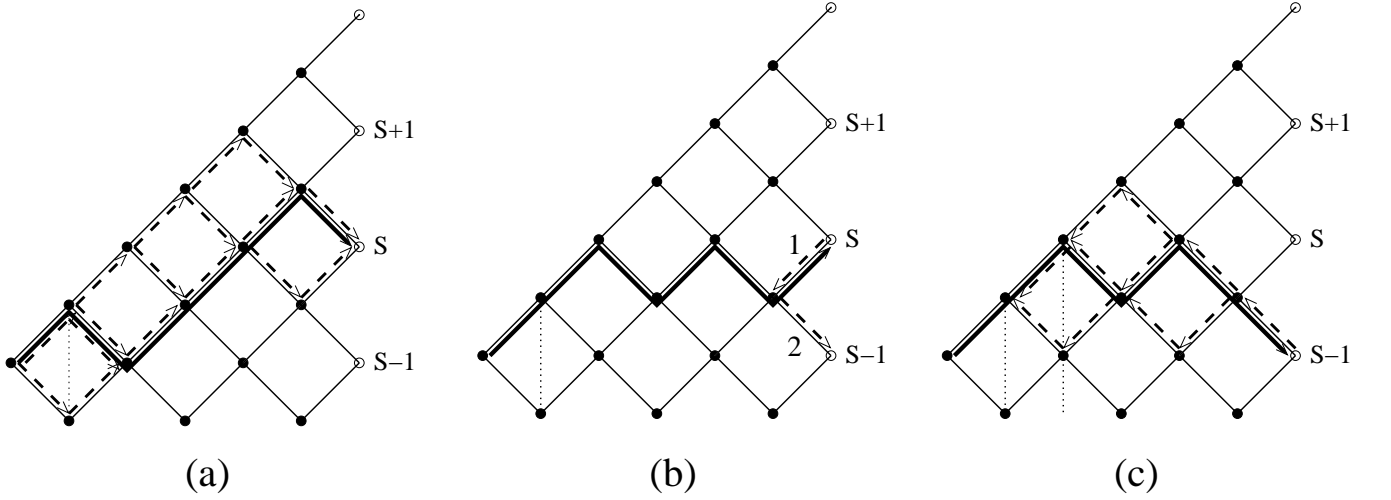}
\caption{An example of spin-tree operations involved in the calculation of the reduced matrix element  $\weck{\gamma' S'}{A_{\mu\nu}^{\Sigma=1}}{\gamma S}$ for case (i) and $S'=S-1$. We consider $q=7$ singly-occupied levels (the multiplicity of the states at the internal nodes $k$ is not shown explicitly). In panel (a) the thick solid line represents the initial path $\gamma$. The paths $\bar{\gamma}$ which result from the push-out operation of state $\nu$ (at position 2 in the normal-ordered tree) are obtained by taking any combination of dashed lines that leads to spin $S$ in the terminal node, starting from node 1. Note that there are nine such paths. In panel (b), we show by the thick solid line one of the possible paths $\bar{\gamma}$ from panel (a) and consider the action of the operator $A_{\mu\nu}^{\Sigma=1}$. The action of the annihilation operator in $A_{\mu\nu}^{\Sigma=1}$ can only result in a backward motion on path $\bar{\gamma}$ (labelled by 1), and leads to intermediate spin $S''=S-\frac{1}{2}$. The action of the creation operator in $A_{\mu\nu}^{\Sigma=1}$ can lead to both $S'=S-1$ and $S'=S$. We choose the path (labelled by 2) that leads to the desired final state $S'=S-1$. Finally, in panel (c) we consider the push-in operation to move $\mu$ back to its position 3 in the normal-ordered representation (denoted by the second dotted vertical line). The paths generated through this procedure can be obtained by any combination of the dashed lines that starts from the terminal node at $S$ and connects to the thick solid line at position 2. There are five such paths.}
\label{figtreeex1}
\end{figure}

Fig.~\ref{figtreeex1} illustrates the spin-tree operations which enter the calculation of $\weck{\gamma' S'}{A_{\mu\nu}^\Sigma}{\gamma S}$. The initial path $\gamma=(S_1,S_2,\ldots,S_{q}=S)$ in a normal tree that connects the root to the desired terminal point (e.g., the thick solid line in Fig.~\ref{figtreeex1}(a)). 
The possible paths $\bar{\gamma}=(\bar{S}_1,\bar{S}_2,\ldots,\bar{S}_{q}=S)$ which result when the level $\nu$ (at $\lambda_i=\nu$) is pushed out to the terminal point to form a canonical $\nu$-tree are represented by any combination of the dashed lines in Fig.~\ref{figtreeex1}(a). Note the simple rule that results from the form of the exchange operator $R_{S_k,\bar{S}_k}(S_{k-1},S_{k+1})$. For $k < i$, $\bar{S}_k= S_k$. For $k \ge i$, the (two) possible values of $\bar{S}_k$ are determined by examining the values of $\bar{S}_{k-1}$ and $S_{k+1}$. The Wigner 6j selection rules dictate that $\bar{S}_k = \bar S_{k-1} + \frac{1}{2}$ for $S_{k+1} = \bar{S}_{k-1}+1$, $\bar{S}_k = \bar S_{k-1} - \frac{1}{2}$ for $S_{k+1} = \bar{S}_{k-1}-1$ and $\bar{S}_k = \bar S_{k-1} \pm \frac{1}{2}$ for $S_{k+1} = \bar{S}_{k-1}$. Thus, given the initial path $\gamma$, the set of paths $\bar{\gamma}$ generated by the push-out operation is easily determined. These rules dictate that $\bar{\gamma}$ and $\gamma$ are identical for $k < i$, that they may differ by at most $\Delta S = \pm 1$ in each internal node for $k \ge i$ and that their final spins are identical.  
  
Next we consider the operation of $A_{\mu\nu}^\Sigma$ in the canonical basis, connecting a canonical $\nu$-tree with terminal spin $S$ to a canonical $\mu$-tree with terminal spin  $\bar{S}'=S$, $S\pm 1$ (the particular values of $\bar{S}'$ depend on $\bar{S}_{q-1}$). In Fig.~\ref{figtreeex1}(b) we show by the thick solid line one of the allowed paths $\bar \gamma$ generated in Fig.~\ref{figtreeex1}(a). For this path, the allowed values for $\bar{S}'$ are $S,S-1$. Since we are interested in the final state $S'=S-1$, we have $\bar{S}'=S-1$, $\bar{\gamma}'=(\bar{S}_1,\ldots,\bar{S}_{q-1},S-1)$, and the canonical matrix element of $A_{\mu\nu}^\Sigma$ is given by Eq.~(\ref{eqredmatel1}) with $S''=\bar{S}_{q-1}$ and $S'=S-1$.   Finally, in Fig.~\ref{figtreeex1}(c) we show by dashed lines the resulting possible paths $\gamma'$  after using the push-in operation to transform from the canonical tree $\bar{\gamma}'$ (thick solid line) to the normal tree (for which $\lambda_j=\mu$). 

Thus, starting from an initial state with quantum numbers $(\gamma, S)$, the non-zero matrix elements of $A_{\mu\nu}^\Sigma$ in the normal basis can be found by choosing all sets of allowed paths $\bar{\gamma}$, $\bar{\gamma}'$ and $\gamma'$ (as described above). For each set, we attach the matrix element $R_{S_k,\bar{S}_k}(\bar{S}_{k-1},S_{k+1})$ to each internal node $k \geq i$, the matrix element $R_{\bar{S}_k',S_k'}(\bar{S}_{k-1}',S_{k+1}')$ to each internal node $k\geq j$, and Eq.~(\ref{eqweckthe1}) to the terminal node. For a given final state $(\gamma',S')$, we then sum over all intermediate paths $\bar{\gamma}$ and $\bar{\gamma}'$. In practice, the summation is carried out only over $\bar\gamma$ (or equivalently $\bar\gamma'$). This is because $\bar{\gamma}'$ and  $\bar{\gamma}$ may differ only in the value of the terminal node $S_q$ which does not enter the summation. 

\subsubsection{Transition from a singly-occupied level to a singly-occupied level}

In this case a particle is moved from a singly-occupied level $\nu$ to a singly-occupied level $\mu$. Contrary to the previous case, the number of singly-occupied levels changes from $q$ to $q-2$ (since level $\mu$ becomes doubly occupied), and the final state is represented by a shorter tree.   Since both levels $\nu$ and $\mu$ exist in the spin tree representing the initial state $(\gamma, S)$, there is only one canonical tree defined by $\lambda_{q-1}=\mu$ and $\lambda_q = \nu$. This canonical tree is obtained from the initial normal tree by applying the push-out operation on both singly-occupied levels $\nu$ and $\mu$. Since we want level $\nu$ to be the last coupled orbital, we first perform the push-out of $\mu$ and then the push-out of $\nu$. This process is described by
\be
(\gamma,S) \stackrel{\bm{U}^{\mu}}{\ra} (\bar{\gamma},\bar{S})_\mu \stackrel{\bm{U}^{\nu}}{\ra}  (\bar{\gamma}',\bar{S}')_{\mu\nu} \stackrel{\a{\nu}{}}{\ra} (\gamma'',S'')_\mu  \stackrel{\ad{\mu}{}}{\ra} (\gamma',S') \;,
\ee
where $\gamma''=(\bar{S}'_1,\ldots,\bar{S}'_{q-1})$ and  $S'' = \bar{S}_{q-1}'$.
 The corresponding reduced matrix element is then given by
\begin{equation}
\weck{\gamma' S'}{A_{\mu\nu}^\Sigma}{\gamma S} = \sum_{\bar{\gamma},\bar{\gamma}'} U_{\gamma\bar{\gamma}}^\mu U_{\bar{\gamma}\bar{\gamma}'}^\nu  \, {\weck{\gamma' S'}{A_{\mu\nu}^\Sigma}{\bar{\gamma}' \bar{S}'}_c} \;,
\end{equation}
where $U$ is the push-out operator defined in Eq.~(\ref{eqpushoutU}). The canonical reduced matrix element in this case is found to be
\be
\weck{\gamma' S'}{A_{\mu\nu}^\Sigma}{\bar{\gamma}' \bar{S}'}_c = (-1)^{2\bar{S}'-\Sigma+1} \sqrt{(2\Sigma+1)(2\bar{S}'+1)(2S''+1)} \sixj{\bar{S}'}{S'}{\Sigma}{1/2}{1/2}{S''} 
\;.
\label{eqredmatel2}
\ee

\subsubsection{Transition from a doubly-occupied level to a single-occupied level}

Here a particle is moved from a doubly-occupied level $\nu$ to a singly-occupied level $\mu$. The level $\mu$ exists in the initial tree $(\gamma, S)$ while the level $\nu$ exists in the final tree $(\gamma',S')$. We implement the following process
\be
(\gamma,S) \stackrel{\a{\nu}{}}{\ra} (\gamma'',S'')_\nu \stackrel{\bm{V}^{\nu}}{\ra} (\bar{\gamma},\bar{S}) \stackrel{\bm{U}^{\mu }}{\ra} (\bar{\gamma}',\bar{S}')_\mu \stackrel{\ad{\mu}{}}{\ra} (\gamma',S') \;,
\ee
where $\gamma''=(S_1,\ldots, S_q=S, S_q \pm 1/2)$,  $S''=S \pm 1/2$ and $\bar{\gamma}'=(S_1',\ldots,S_q',\bar{S}_{q+1}')$.  The corresponding reduced matrix element in the normal basis is given by
\begin{equation}
\weck{\gamma' S'}{A_{\mu\nu}^\Sigma}{\gamma S} = \sum_{\gamma'',\bar{\gamma},\bar{\gamma}'} V_{\gamma''\bar{\gamma}}^\nu U_{\bar{\gamma}\bar{\gamma}'}^\mu  \, {\weck{\gamma' S'}{A_{\mu\nu}^\Sigma}{\gamma S}_c} \;,
\end{equation}
and the corresponding canonical reduced matrix element is 
\be
{\weck{\gamma' S'}{(A_{\mu\nu})^\Sigma}{\gamma S}_c} = (-1)^{S+S''-\Sigma+1/2} \sqrt{(2\Sigma+1)}(2S''+1) \sixj{S}{S'}{\Sigma}{1/2}{1/2}{S''} \;.
\label{eqredmatel3}
\ee

\subsubsection{Transition from a doubly-occupied level to an empty level}

Here a particle is moved from a doubly-occupied level $\nu$ to an empty level $\mu$. This generates a canonical $\nu\mu$-tree with $\lambda_{q+1}= \mu$ and $\lambda_{q+2}=\nu$. The normal tree of the final state is then obtained by push-in operations on both levels $\nu$ and $\mu$ (the final state has two more singly-occupied levels than the initial state).  We thus implement the following sequence
\be
(\gamma,S) \stackrel{\a{\nu}{}}{\ra} (\gamma'',S'')_\nu \stackrel{\ad{\mu}{}}{\ra} (\bar{\gamma}',\bar{S}')_{\nu\mu} \stackrel{\bm{V}^{\mu}}{\ra} (\bar{\gamma},\bar{S})_\nu \stackrel{\bm{V}^{\nu }}{\ra} (\gamma',S') \;,
\ee 
where $\gamma''=(S_1,\ldots, S_q=S, S_q \pm 1/2)$, $S''=S \pm 1/2$ and $\bar{\gamma}'=(S_1,\ldots, S_q, S'', S'' \pm 1/2)$. It follows that the allowed values of $\bar{S}'$ are $S,S\pm 1$. The reduced matrix element can then be expressed in the following form
\begin{equation}
\weck{\gamma' S'}{A_{\mu\nu}^\Sigma}{\gamma S} = \sum_{\gamma'',\bar{\gamma},\bar{\gamma}'}  V_{\bar{\gamma}'\bar{\gamma}}^\mu V_{\bar{\gamma}\gamma'}^\nu \, {\weck{\bar{\gamma}' \bar{S}'}{A_{\mu\nu}^\Sigma}{\gamma S}_c} \;,
\end{equation}
and the corresponding canonical reduced matrix element is given by
\be
{\weck{\bar{\gamma}' \bar{S}'}{A_{\mu\nu}^\Sigma}{\gamma S}_c} = (-1)^{\bar{S}'+S-\Sigma+1} \sqrt{(2\Sigma+1)(2\bar{S}'+1)(2S''+1)} \sixj{S}{\bar{S}'}{\Sigma}{1/2}{1/2}{S''} \;.
\label{eqredmatel4}
\ee

\section{Other technical aspects}

\subsection{Organization and truncation of many-particle states in the good-spin basis}

In practical calculations it is often necessary to truncate the many-particle Hilbert space. For that purpose the many-particle good-spin states are arranged in ascending order of their total energy in the universal Hamiltonian (including the exchange interaction).  Below we describe briefly the implementation of an enumeration scheme which takes maximal advantage of the spin and fermionic structure of the many-body Hilbert space.

The universal Hamiltonian eigenstates are labelled by  $\bra{ \bm{n} \gamma S M}$. We first enumerate these states according to their ``distance'' from the Fermi sea as measured by the electron-hole excitation metric.  Thus,
\begin{equation}
\bm{n} \rightarrow \bra{\bm{h_2};\bm{h};\bm{e_2};\bm{e}} \;,
\end{equation}
where $\bm{h}$ is a vector of occupied hole states, $\bm{h_2}$ is a vector of doubly-occupied hole states, and $\bm{e}$, $\bm{e_2}$ describe similarly the electron states.  
Next we implement the spin-labelling scheme. Since the doubly-occupied levels have zero spin, the spin structure is completely determined the singly-occupied levels. This structure is best represented by a spin-tree $\gamma=(S_1,S_2,\ldots,S)$. The resulting good-spin eigenstates of the universal Hamiltonian are then arranged in ascending order of their energy $E_{\bm{n} \gamma S}$ and truncation can be done by choosing a suitable energy cutoff. This scheme allows for inclusion of low-energy large-$S$  states (in the case of ferromagnetic coupling) already at the level of the unperturbed basis states. This is particularly advantageous near the Stoner instability, i.e., for $J_s \sim \Delta$.  

Convergence in a truncated basis can be improved by using banded matrices to describe the spin-orbit matrices $\Gamma^\perp$ and $\Gamma^\parallel$ in Eq.~(\ref{spin-orbit}).  In particular, we have observed better convergence for Gaussian-banded matrices.  Banded random matrices with a band width of order $\sqrt{N}$ ($N$ is the dimension of the matrix) have  similar statistics as in standard RMT. 

\subsection{Spin-dependent observables}

The advantage of using a good-spin basis is transparent in the computation of spin-dependent observables. Physical observables can be classified as tensor operators under spin rotations and it is sufficient to calculate their reduced matrix elements, irrespective of the magnetic quantum numbers. We have already used this in the calculation of the matrix elements of the spin-orbit interaction. 

Spin-orbit interaction breaks spin symmetry, and in particular the ground state does not have good spin. However, it is still of interest to find the ground-state spin distribution. To find this distribution it is necessary to expand the ground-state in a good-spin basis. In our approach, such an expansion is automatically obtained once the many-body Hamiltonian is diagonalized in the good-spin basis.  

Another interesting spin-dependent observable is the transverse spin correlator
\begin{equation}
S_+(\om) = 
-\frac{1}{\pi} \im{\langle \hat{S}_+(\om) \hat{S}_-(-\om) \rangle} \;.
\label{eqspincorran1}
\end{equation}
This quantity can be easily computed by expanding the ground-state $|0 \rangle$ and excited states $|I\rangle$ in the good-spin basis. At zero temperature, we have
\begin{equation}
S_+(\om) = \sum_{I}  \left| \sum_{\bm{n}\gamma S M} \sqrt{S(S+1)-M(M+1)}   \braket{I}{\bm{n}\gamma S M+1} \braket{\bm{n}\gamma SM}{0} \right|^2 \delta (\om - E_I) \;,
\end{equation}
where we have used the known matrix elements of $\hat S_+$ in the good-spin basis 
\begin{equation}
\ket{\bm{n}'\gamma' S'M'} \hat{S}_+ \bra{\bm{n}\gamma SM} = \sqrt{S(S+1) - M(M+1)} \, \delta_{M',M+1}\delta_{S',S}\delta_{\gamma',\gamma}\delta_{\bm{n}'\bm{n}} \;.
\end{equation}

\section{Conclusion}

The problem of spin-orbit scattering in the presence of exchange correlations does not have a closed solution in the crossover between different symmetries of the spin-orbit scattering problem.  We have developed an approach to solve the problem numerically that is based on an unconventional choice of the many-particle basis. Rather than using a basis of Slater determinants built from the single-particle wave functions of the single-particle Hamiltonian, we use a correlated good-spin basis. This basis is composed of the eigenstates of the universal Hamiltonian which includes an exchange interaction in the absence of spin-orbit scattering. For a given set of occupation numbers of the spin-degenerate orbital levels, we construct many-particle states of good total spin by coupling the 
singly-occupied levels as spin-$1/2$ particles.  To distinguish between states with same spin, we use spin-tree diagrams. 

The spin-orbit interaction terms are one-body operators, and they can be decomposed into sum of spherical tensors under spin rotations. We have used angular momentum algebra to calculate the reduced matrix elements of these tensor operators in closed form. In a basis that is canonical with respect to the transition operator, the reduced matrix elements can be expressed in terms of Wigner 6j symbols. To find the reduced matrix elements in the normal basis, it is necessary to make a unitary transformation from a normal tree to a canonical tree. The calculation of this unitary transformation involves a series of spin exchanges, and each such exchange can also be expressed in terms of a corresponding 6j symbol.

The use of the good-spin basis has several advantages:
  
  \begin{itemize}
  
  \item
  It is only necessary to calculate reduced matrix elements of the spin-orbit terms. Such reduced matrix elements are independent of the magnetic quantum numbers, and the actual matrix elements are easily calculated using the Wigner-Eckart theorem. Furthermore, both the ${\bf a}_\perp$ and ${\bf a}_\parallel$ terms of the spin-orbit interaction are calculated from the same reduced matrix elements of the operators $A^{\Sigma=1}_{\mu\nu}$.  The ${\bf a}_\perp$ and ${\bf a}_\parallel$ terms are described, respectively, by the $m=0$ and $m=\pm 1$ components of $A^{\Sigma=1}_{\mu\nu}$. The dependence of the (non-reduced) matrix elements on the specific $m$ component of  $A^{\Sigma=1}_{\mu\nu}$ enters only through a corresponding Clebsch-Gordan coefficient. 

\item
The exchange correlation energy is fully included in the unperturbed basis of the universal Hamiltonian. Such a basis is particularly useful when the exchange coupling is ferromagnetic and near the Stoner instability.  If an energy-cutoff truncation is necessary (e.g., in large model spaces), our truncated good-spin basis will include high spin states that lie low in energy because of the attractive exchange interaction.

\item
Spin-dependent observables such as the ground-state spin distribution and spin excitation function are easily calculated since the calculated eigenstates are already expressed as a superposition of good-spin states.

\item
Our approach is not limited to the spin-orbit coupling problem but is applicable for any perturbation that is added to the universal Hamiltonian. Any one-body operator can be decomposed into spin tensor operators $A^\Sigma_{\mu\nu}$ with $\Sigma=0,1$, and the matrix elements of such operators have already been calculated here in a closed form. We can also express any two-body perturbation as a superposition of tensor products of two such one-body tensor operators $A^{\Sigma}_{\mu\nu}$ and $A^{\Sigma'}_{\mu'\nu'}$. 

\end{itemize}

In this work, we have presented the good-spin basis formalism and its technical aspects. This formalism is useful in the study of problems involving exchange correlations in quantum dots. One such problem is the spin-orbit scattering problem discussed in this manuscript. Ground-state spin distribution, spin excitation function and conductance peak statistics in almost-isolated dots can be studied as a function of the exchange constant $J_s$ and the scaled spin-orbit crossover parameters $x_\perp$ and $x_\parallel$. In particular, it would be interesting to compare with signatures of quantum critical 
fluctuations~\cite{murthy04a} that were calculated in the crossover between various symmetry limits of the spin-orbit scattering problem.

\section*{Acknowledgments}

We thank A. Douglas Stone, Sebastian Schmidt and Stefan Rotter for useful discussions. This work was supported in part by the U.S. Department of Energy under Grant DE-FG-02-91ER40608.

\section*{Appendix: Derivation of the reduced matrix element of $A_{\mu\nu}^\Sigma$ in the canonical basis}

Here we derive the reduced matrix element 
${\weck{\bar\gamma' S'}{A_{\mu\nu}^\Sigma}{\bar\gamma S}_c}$ in the canonical basis for case (i), describing the transition from a singly-occupied level to an empty level (see Sec. \ref{case1}). To this end, we will compute the matrix element ${_\mu\ket{\gamma' S'M'}(A_{\mu\nu})^\Sigma_m \bra{\gamma SM}_\nu}$. Using the expression for the irreducible tensor operator $(A_{\mu\nu})^\Sigma_m$ in Eq.~(\ref{eqirrTOgen}) and the resolution of the identity, $\sum_{\bar\gamma'' S'' M''} \bra{\bar\gamma'' S''M''}\ket{\bar\gamma''S''M''}
 =1$, we have
\begin{align}
_\mu \ket{\bar\gamma' S'M'} \left( \ad{\mu}{} \otimes \tilde{a}_{\nu}{} \right)^{\Sigma}_{m} \bra{\bar\gamma SM}_\nu = & \sum_{\sigma\sigma'M''} (-1)^{(1-\sigma')/2} \cg{1/2,\sigma/2}{1/2,-\sigma'/2}{\Sigma,m}  \nonumber \\
& 
 _\mu \ket{\gamma' S'M'} \ad{\mu}{\sigma} \bra{\bar\gamma'' S''M''}\ket{\bar\gamma''S''M''}a_{\nu\sigma'} \bra{\bar\gamma SM}_\nu \;,
\label{eqRedMatcase11}
\end{align}
where $\bar\gamma''=(S_1=1/2,\bar S_2,\ldots,\bar S_{q-1})$ is a normal tree, and $S''=\bar S_{q-1}$ is the intermediate spin after a particle is removed from orbital $\nu$.
Here, $n_\mu= n_\nu=0$ for $\bra{\bar\gamma'' S''M''}$, while $n_\mu=0$, $n_\nu=1$ for $\bra{\bar\gamma SM}_\nu$  and $n_\mu=1$, $n_\nu=0$ for $\bra{\bar\gamma' S'M'}_\mu$. Since $\ad{\mu}{\sigma}$ transforms as an irreducible tensor operator of rank $1/2$ under spin rotations, we have
\begin{equation}
{_\mu\ket{\bar\gamma' S'M'}} \ad{\mu}{\sigma} \bra{\bar\gamma'' S''M''} = (-1)^{q-1}\, \cg{S'',M''}{1/2,\sigma /2}{S'M'}  \;,
\end{equation}
and
\begin{equation}
\ket{\bar\gamma''S''M''}a_{\nu\sigma'} \bra{\bar\gamma SM}_\nu = {_\nu\ket{\bar\gamma SM}} \ad{\nu}{\sigma'} \bra{\bar\gamma''S''M''}^* = (-1)^{q-1} \, \cg{S'',M''}{1/2,\sigma'/2}{SM}  \;.
\end{equation}
The factors $(-1)^{q-1}$ account for the fermionic nature of the spin operators. Substituting these relations in Eq.~(\ref{eqRedMatcase11}) 
\begin{align}
_\mu \ket{\gamma' S'M'} \left( \ad{\mu}{} \otimes \tilde{a}_{\nu}{} \right)^{\Sigma}_{m} \bra{\gamma SM}_\nu = & \sum_{\sigma\sigma'M''} (-1)^{(1-\sigma')/2} \cg{1/2,\sigma/2}{1/2,-\sigma'/2}{\Sigma,m}  \nonumber \\
& 
 \times \cg{S'',M''}{1/2,\sigma/2}{S'M'} \cg{S'',M''}{1/2,\sigma'/2}{SM} \;.
\label{eqRedMatcase12}
\end{align}
This can be rewritten in terms of Wigner 3j symbols as 
\begin{align}
_\mu \ket{\gamma' S'M'} \left( \ad{\mu}{} \otimes \tilde{a}_{\nu}{} \right)^{\Sigma}_{m} \bra{\gamma SM}_\nu = & \sum_{\sigma\sigma'M''} (-1)^{S + S' - S'' - \frac{1}{2} - M'} (-1)^{S''+ \frac{1}{2}\sigma - M} \sqrt{(2\Sigma+1)(2S+1)(2S'+1)} \nonumber \\ & \times \threej{\frac{1}{2}}{\frac{1}{2}}{\Sigma}{\frac{1}{2}\sigma}{-\frac{1}{2}\sigma'}{-m} 
\threej{\frac{1}{2}}{S''}{S}{\frac{1}{2}\sigma'}{M''}{-M}
\threej{S''}{\frac{1}{2}}{S'}{-M''}{-\frac{1}{2}\sigma}{M'} \;.
\label{RedMat-3j}
\end{align}
Using a sum rule relating 3j symbols to a 6j symbol~\cite{messiah_bookv2} and symmetry properties of the 3j symbol, Eq.~(\ref{RedMat-3j}) can be expressed as 
\begin{align}
_\mu \ket{\gamma' S'M'} \left( \ad{\mu}{} \otimes \tilde{a}_{\nu}{} \right)^{\Sigma}_{m} \bra{\gamma SM}_\nu = &
(-1)^{- S' - S'' - \frac{1}{2} - \Sigma} \sqrt{(2\Sigma+1)(2S+1)(2S'+1)} \nonumber \\
& \times (-1)^{S'-M'} \threej{S'}{\Sigma}{S}{-M'}{m}{M}
\sixj{S}{S'}{\Sigma}{\frac{1}{2}}{\frac{1}{2}}{S''} \;.
\end{align}
Comparing with Wigner-Eckart theorem, Eq.~(\ref{eqweckthe1}), we obtain Eq.~(\ref{eqredmatel1}) for the reduced matrix element.


\end{document}